\documentstyle[preprint,eqsecnum,aps,epsfig]{revtex}

\begin{document}
\draft
\title{Orbital Magnetic Dipole Mode in Deformed Clusters: A Fully Microscopic
Analysis}
\author{V. O. Nesterenko$^{1,2}$, W. Kleinig$^{2,3}$, F. F. de Souza Cruz$^{1}$ and
N. Lo Iudice$^{4}$}
\address{$^{1}$Departamento de Fisica, Universidade Federal de Santa Catarina,
88040-900, Florianopolis, SC, Brasil\\
$^{2}$ Bogoliubov Laboratory of Theoretical Physics, Joint Institute for
Nuclear Research, 141980 Dubna, Moscow region, Russia\\
$^{3}$Technische Universit\"{a}t Dresden, Institut f\"{u}r Analysis, D-01062
Dresden, Germany\\
$^{4}$Universit${\grave{a}}$ di Napoli ''Federico II'' and Istituto\\
Nazionale di Fisica Nucleare, Dipartamento di Scienze Fisiche, Monte S.
Angelo, Via Cintia I-80126, Napoli, Italy}
\date{\today}
\maketitle

\begin{abstract}
The orbital M1 collective mode predicted for deformed clusters in a
schematic model is studied in a self-consistent random-phase-approximation
approach which fully exploits the shell structure of the clusters. The
microscopic mechanism of the excitation is clarified and the close
correlation with E2 mode established. The study shows that the M1 strength
of the mode is fragmented over a large energy interval. In spite of that,
the fraction remaining at low energy, well below the overwhelming dipole
plasmon resonance, is comparable to the strength predicted in the
schematic model. The importance of this result in view of future experiments
is stressed.  
\end{abstract}


\preprint{HEP/123-qed}

\narrowtext

Among the collective excitations which may occur in metal clusters, the
magnetic dipole mode predicted for deformed clusters in a schematic model 
\cite{LiSt} has unique and appealing properties which deserve a deeper
investigation. This excitation, which is the analogue of the scissors mode
predicted \cite{IuPa-78} and observed \cite{ Bohle84} in deformed nuclei, is
promoted by a rotational oscillations of the valence electrons against the
jellium background. Indeed, in the semiclassical approach \cite{LiSt}, the
displacement field of the mode is composed of a rigid rotational velocity
field plus a quadrupole term which comes from the boundary condition that
the velocity flow vanishes on the deformed surface. The distortion of the
momentum Fermi sphere generates a restoring force of the rotational
oscillations. The mode is characterized by the magnetic quantum number $%
K^{\pi }=1^{+}$ and falls at an excitation energy \cite{LiSt} 
\begin{equation}
\omega _{M1}=\sqrt{2}\omega _{0}\delta \frac{1}{\sqrt{1+5\omega
_{0}^{2}/\omega _{p}^{2}}}\ \simeq \frac{20.7}{r_{s}^{2}}N_{e}^{-1/3}\delta
\ eV  \label{omscis}
\end{equation}
where $\omega _{0}=(2\epsilon _{F}/mr_{s}^{2})^{1/2}N_{e}^{-1/3}$ and $%
\omega _{p}$ are respectively the harmonic oscillator (HO) and the plasma
frequencies, $\delta $ is the deformation parameter, $\ r_{s}$ is the
Wigner-Seitz radius, $\epsilon _{F}$ is the Fermi energy ( $r_{s}=2.1\AA $
and $\epsilon _{F}=3.1eV$ for Na clusters), and $N_{e}$ is the number of
valence electrons in a cluster. The latter is related to the number $N$ of
atoms in a cluster by $N=N_{e}\ $or $N=N_{e}+1$ according that the cluster
is neutral or has a positive charge $Z=+1$. The mode gets a M1 strength
given by 
\begin{eqnarray}
B(M1) &=&2\mid \langle K^{\pi }=1^{+}\mid \sum_{i}l_{x}(i)\mid 0\rangle \mid
^{2}\mu _{b}^{2}  \nonumber \\
&=&\Im \omega _{M1}\mu _{b}^{2}\ \ \simeq N_{e}^{4/3}\delta \ \mu _{b}^{2}
\label{B(M1)}
\end{eqnarray}
where $\Im =2/3$ $N_{e}m<r^{2}>$ = $2/5$ $r_{s}^{2}mN_{e}^{5/3}$ is the
collective mass parameter. As the formulas show, the M1 mode is peculiar of
deformed clusters. Its occurrence would represent a unique and unambiguous
fingerprint for the onset of quadrupole deformation. The main indicator of
the deformation available so far is the splitting of the E1 resonance which,
however, is often washed out or not properly resolved experimentally. The
energy formula (1) reveals another appealing property. The M1 mode falls
well below the energy of the overwhelming E1 resonance and, therefore, has
good chances of being detected experimentally. In view of such a
possibility, it is of the utmost importance to test the predictions of the
schematic model by carrying out a microscopic calculation which fully
exploits the shell structure of the clusters. Such a calculation should shed
light on the microscopic mechanism which generates the mode and should
reveal, eventually, new properties connected with the shell structure It
certainly will ascertain if and to what extent the M1 strength is fragmented
and quenched. Clearly such a mode can be detected and can be used as a
signature for deformation only if its M1 strength remains concentrated in a
reasonably narrow energy range.

We performed our calculation in a self-consistent RPA (SRPA) approach \cite
{Ne-PRA,Ne-Tsu,Kle-EPJ} which in the most general formulation\cite{Ne-PRA}
is based on the Kohn-Sham functional \cite{KS}. Here we skipped the
self-consistent derivation of the one-body potential and adopted the
phenomenological deformed Woods-Saxon well. On the other hand, we determined
self-consistently the two-body potential starting with a  set of
displacement fields ${\bf \nabla }f_{L21}$, where 
\begin{equation}
\ f_{L21}=r^{L}(Y_{21}+Y_{21}^{*}),\qquad L=2,4,6,8.  \label{Fields}
\end{equation}
The resulting interaction was a sum of weighted separable terms 
peaked on different slices of the system. Due to its close link to the detailed structure of the
system, such an  interaction, in spite of its separable form,
came out to be quite suitable for describing its dynamical properties.
The SRPA fully exploits the shell structure and, as shown for
the dipole response\cite{Kle-EPJ}, reaches the accuracy of the most refined
and complete RPA approaches. Moreover, it
preserves the simplicity of the schematic model\cite{LiSt}
which can be easily recovered if an anisotropic HO potential plus the single operator $f_{221}=$ $
r^{2}(Y_{21}+Y_{21}^{*})$ are used.

The choice of a quadrupole-like field was motivated by its close connection
with the generator of the rotational oscillation, namely the angular
momentum (see \cite{I98} and Refs. therein). Such a link can be easily
established for $L=2$ in the HO space , where one finds 
\begin{equation}
\langle p\mid l_{x}\mid h\rangle \simeq \sqrt{\frac{\ 4\pi }{15}}mb(\omega
_{0)}\langle p\mid f_{221}\mid h\rangle   \label{LQequiv}
\end{equation}
being $b(\omega _{0})=2\omega _{0}$ for the $\Delta {\cal N}=0$ space and $%
b(\omega _{0})=\delta \omega _{0}$ for $\Delta {\cal N}=2$ (${\cal N}$ is a
principle shell quantum number). On the other hand, the quadrupole fields
act obviously also in the E2 channel. Consistency requires that both,
magnetic dipole and electric quadrupole, excitations should be treated
contextually. It is worth noting that, the choice made for our displacement
fields (see Eq.\ref{Fields}) is quite general for our purposes. Indeed,
since spin-orbit coupling in clusters is negligible, the orbital excitations
are decoupled from the spin ones, so that the spin-spin interaction can be
safely neglected. The spin-quadrupole fields can be also ignored. In the
nuclear systems, they are known to affect only the spin channel by
renormalizing the spin-spin interaction.

The parameters of the Woods-Saxon potential $V_{WS}=V_{0}/[1+exp{\bf (}R(\Theta
)-r)/a_{0}{\bf )}]$ with $R(\Theta )\ =R_{0}(\beta _{0}+\beta _{2}Y_{20}(\Theta ))$
and $R_{0}=r_{0}N^{1/3}$ were adjusted so as to reproduce the Kohn-Sham+SRPA
results for the dipole plasmon in spherical sodium clusters \cite{Ne-PRA}.
The fit yielded $r_{0}=2.5\AA $, $V_{0}=-7.2$ eV and $a_{0}=1.25\AA $ for
singly charged clusters and $r_{0}=2.4\AA $, $V_{0}=-5.7$ eV and $%
a_{0}=1.11\AA $ for neutral clusters. The values of the deformation
parameter $\delta =\sqrt{45/16\pi }\beta _{2}$ were extracted from the
experimental data \cite{SH-exp} (for $N_{e}\le 34$) following the
prescription of Ref. \cite{LiSt}, or were taken from the calculations \cite
{FP96} (for $N_{e}>34$). Only clusters with measured or predicted {\it axial}
quadrupole deformation were considered. Equal deformation parameters were
used for both charged and neutral clusters. As shown in Ref.\cite{Ne-Tsu},
the SRPA calculations with these parameters account well for the observed
deformation splitting of the dipole plasmon in deformed Na clusters.

The most meaningful results of the calculation are presented in Figs. 1-3.
In Fig. 1 the M1 strength distribution is plotted for singly charged
clusters varying from $N=15$ to $295$. In order to simulate the temperature
broadening, we smoothed out the M1 strength with the Lorentz weight using
the averaging parameter $\Delta =0.05$ eV. 
Such a simulation is in general rather rough as compared to an
explicit treatment of electronic and ionic thermal fluctuations.
Nonetheless, since we do not pretend to describe specific thermal
effects, this simulation should be sufficient  for our purposes, 
if we confine ourselves within the temperature interval 300-600 K, 
where jellium aproximation is appropriate.
We also gave in Fig. 1 a quantitative
estimate of the Landau damping by computing the width $\Gamma $ of the
resonance which ideally envelops all the peaks above a threshold value fixed
to be one-half the height of the highest peak \cite{Kle-EPJ}. This
definition yields the standard full width at half maximum (FWHM) in the
simplest case of one-peak structure. The plot shows that, as the size of the
cluster increases, the whole M1 strength is shifted downward with rising
magnitude and fragmentation. Only in going from the light prolate $%
Na_{27}^{+}$ to the light oblate $Na_{35}^{+},$ this trend is not observed.
In heavy clusters with $N_{e}\sim 300,$ the M1 strength reaches the huge
values 350-400 $\mu _{b}^{2}$. The fragmentation (Landau damping) gets also
very pronounced, since $\Gamma $ and $\bar{\omega}$ become comparable. Due
to the small value of $\bar{\omega}$, however, the strength remains
concentrated in a rather narrow energy interval.

The softening of the mode as well as the enhancement of the M1 strength can
be nicely explained within the semiclassical model with the decreasing
importance of the surface with respect to the bulk as the sizes of the
cluster increase. This causes a faster increase of the mass parameter with
respect to the restoring force constant coming almost entirely from a
surface shear, with consequent lowering of the energy centroid (\ref{omscis}%
)and enhancement of the M1 strength (\ref{B(M1)}). A more detailed and
exhaustive explanation is provided by the microscopic excitation mechanism.
By expanding the deformed single-particle wave function into a spherical
basis $\mid m\rangle =\sum_{nl}a_{nl}^{m}\mid nlm\rangle $ and accounting
for the fact that each $\mid m\rangle $ state is, in general, dominated by a
single spherical configuration $\mid nlm\rangle $, one obtains the
transition amplitude 
\begin{equation}
\ \langle \ m^{\prime }\mid {\hat{l}}_{\pm }\mid m\ \rangle \ \simeq \mp
\delta _{m^{\prime },m\pm 1}\ \sqrt{l(l+1)-m(m\pm 1)}.  \label{lmatr}
\end{equation}
Clearly, the main contribution to the transition amplitude comes from orbits
with high angular momentum $l$ and small magnetic quantum number $m.$ On the
other hand, orbits with high $l$ values are present only in heavy clusters.
Hence the enhancement of the M1 strength. At the same time, as the sizes of
the cluster increase with consequent increment of the number of high values
of $l$, the density of the particle-hole $\left( p-h\right) $ levels
increases and their relative spacings decrease, causing an overall downward
shift and a more pronounced fragmentation of the M1 strength. The above
formula enables to sharpen the geometrical picture of the mode. Since most
of the strength comes from orbits with high $l$ and small $m$ values, it
follows that the oscillatory rotational motion is promoted mainly by the
orbits which are almost orthogonal to the equatorial plane.

In light clusters the M1 transition is promoted mainly by one or two
configurations. Indeed, the left and right peaks are due by more than $95\%$
to the p-h components [200]-[211] and [202]-[211] in $Na_{15}^{+}$,
[312]-[321] and [310]-[321] in $Na_{27}^{+},$ and [321]-[310] and
[321]-[312] in $Na_{35}^{+}$, having adopted the Nilsson-Clemenger notation $%
Nn_{z}\Lambda $\cite{C85} for the single-particle orbitals. The reason of
the small $p-h$ admixture induced by the residual interaction is simple. The 
$p-h$ configurations are very few and far apart in energy. The only
observable effect of the interaction is therefore a shift of the M1
strength. We may therefore conclude that in light nuclei the M1 mode has the
character of a single-particle excitation. Only in heavy clusters the
collective nature of the mode appears evident.

Fig. 2 shows the M1 and E2 responses of $Na_{119}^{+}$ over a much wider
energy interval. For a more homogeneous comparison, we give the
photoabsorption cross sections, $\sigma (M1,\mu =1)\sim \sum B(M1,\mu
=1)\omega $ and $\sigma (E2,\mu =1)\sim \sum B(E2,\mu =1)\omega ^{3},$
rather than the strengths. The M1 spectrum is composed of several, roughly
equally spaced, resonances, enveloping closely packed transitions, coming
respectively from $\Delta {\cal N=}$0, 2, 4, ... $p-h$ excitations. The
group of $\Delta {\cal N=}$0 transitions correspond to the low-lying M1 mode
predicted in the schematic model\cite{LiSt}. The others have no classical
counterpart. This large scale fragmentation limits drastically the extent of
validity of approaches which rely entirely on sum rules. It is worth noting,
on the other hand, that the cross section, being proportional to the energy
weighted M1 strength, magnifies the high energy transitions. Had we plotted
the M1 strength, we would have observed a most prominent peak positioned in
the lowest energy region and several others, much less pronounced, at higher
energy. It is also to be pointed out that the high energy peaks which are
physically relevant, namely the ones below the ionization threshold (which
is 3.8 eV in light clusters and 3.2-3.4 eV in the heavier ones), overlap
mostly with the dipole plasmon resonance and, therefore, are hardly
detectable.

The lower panel of Fig. 2 shows that, consistently with the HO relation (\ref
{LQequiv}), the E2 strength covers the same energy regions of the M1
strength. It is, however, dominant over the M1 transition only in the
intermediate region, which is in any case the domain of the dipole plasmon
resonance, and almost absent in the low-energy region. This latter interval
is {\sl exclusively} covered by the M1 mode, consistently with the
predictions of the schematic model.

A more quantitative comparison with this model is presented for the low
energy mode in Fig. 3. Although the schematic Eqs. (1)-(2) were derived for
neutral clusters only, we considered both charged and neutral clusters.
Indeed, the results change very little in moving from one kind to the other.
The energy centroids of the low-energy M1 transitions scale with deformation
and the number of valence electrons basically according to the law derived
in the schematic model \cite{LiSt}. The summed M1 strength scales according
to the schematic law only in heavy clusters, but fluctuates strongly in the
light ones. These fluctuations reflect the single-particle nature of the
transitions and, in principle, invalidate the schematic model for clusters
of these sizes. The low-lying M1 strength is of order of  the semiclassical 
estimates  and even larger in light clusters. Moreover, this strength summed in the interval 1-6 eV exceeds systematically (up to 10-100$\%$) the estimate (2). This looks surprising since in the schematic model the mode gets the total M1 strengh by construction.
This apparent paradox is solved if we
recall that the semiclassical calculation of the strength is fully
equivalent to its RPA evaluation in the $\Delta {\cal N}=0$ HO space. Due to
the degeneracy of the $l$ configurations in the $\Delta {\cal N}=0$ HO
space, a given state $\mid Nn_{z}\Lambda \rangle
=\sum_{l}a_{Nl}^{n_{z}\Lambda }\mid Nl\Lambda \rangle $ is not dominated by
a single configuration $\mid Nl\Lambda \rangle ,$ 
but involves contributions of all orbits with comparable amplitudes $%
a_{Nl}^{n_{z}\Lambda }.$ The resulting M1 transition amplitude does not get
its main contribution from the orbits with largest angular momentum, as in
our case, but is an algebraic weighted sum of different contributions (with
small weights) from all configurations, with both large and small $l$
values. Hence the enhanced M1 strength produced by our calculations.

The main results of our RPA calculation are: i) The M1 strength, at least in
heavy clusters, is not concentrated only at low energy, as predicted by the
schematic model, but spreads over a large energy region among equally spaced
peaks corresponding to $\Delta {\cal N}=0,2,4\cdots $ $p-h$ transitions. ii)
In spite of that, closely packed M1 transitions still fall at low energy and
carry an overall strength which is comparable to the value predicted in
the schematic collective model. Such a strength can become huge in heavy
clusters. It can reach the impressive value of 350-400 $\mu _{b}^{2}$
already at $N_{e}\sim 300.$ iii) The energy centroid of these transitions
scales with the deformation and the number of valence electrons as in the
schematic model only in heavy clusters. In the light ones, the strong
fluctuations of the summed M1 strength invalidate the schematic model.
iv)The crucial role of the quadrupole field in promoting the M1 mode is
confirmed by the close correlation established quantitatively between M1 and
E2 modes. While however the E2 strength is concentrated mostly in the
uninteresting region covered by the plasmon dipole resonance, the M1 is the
only dominant mode at low energy with a strength which becomes huge in heavy
clusters. These properties render the mode quite
accessible to experiments. Its occurrence not only would
indicate the onset of deformation but would enable to measure the
deformation itself by exploiting the scaling properties of the centroids
with deformation and with the sizes of the clusters.

\newpage
Figure captions.

Figure 1. Energy distribution of the M1 strength over the interval 0-1 eV
for sodium clusters ranging from N=15 to 295. The deformation parameter $%
\delta $, the energy centroid $\bar{\omega}$, the quantity $\Gamma $ for
estimating the Landau damping, and the summed M1 strength ($\sum B(M1)$) are
given for each cluster. 

Figure 2. Plot of the M1 and E2 SRPA photoabsorption cross-sections over the
full energy range in $Na_{119}^{+}$. The  curves give the Lorentz averaged
SRPA response, while the underlying bars show the pure discrete spectra,
which better illustrate the Landau damping. 

Figure 3. Ratios between SRPA energy centroids (top) and M1 strengths
(bottom), summed over 0-1 eV, and the corresponding schematic estimates
(Eqs.1-2) for charged (stars) and neutral (triangles) clusters with 
$N_{e}=$14,18,26,34,118,278, and 294.


\begin{references}
\bibitem{LiSt}  
E. Lipparini and S. Stringari, Phys.\ Rev.\ Lett. {\bf 63}, 570 (1989); Z.\
Phys. D {\bf 18} 193 (1991).

\bibitem{IuPa-78}  
N. Lo Iudice and F. Palumbo, Phys.\ Rev.\ Lett. {\bf 41}, 1532 (1978).

\bibitem{ Bohle84}  
D. Bohle {\sl et al.}, Phys. Lett. {\bf B137, }27 (1984).

\bibitem{Ne-PRA}  
V. O. Nesterenko, W. Kleinig, V. V. Gudkov, N. Lo Iudice and J. Kvasil,
Phys.\ Rev. A {\bf 56}, 607 (1997).

\bibitem{Ne-Tsu}  
V. O. Nesterenko and W. Kleinig, Proc. Int. Symp. {\it Similar. and Differ.
between Atom. Nucl. and Clust.} (Tsukuba, Japan, 1997), AIP Conf. Proc. 416,
Woodbury, New York, ed. Y. Abe, I. Arai, S.M. Lee and K. Yabana, p.77 (1998).

\bibitem{Kle-EPJ}  
W. Kleinig, V. O. Nesterenko, P. -G. Reinhard, and Ll. Serra, Eur.\ Phys.\
J.\ D{\bf 4}, 343 (1998).

\bibitem{KS}  
W.Kohn and L.J.Sham, Phys.\ Rev. {\bf 140}, A1133 (1965).

\bibitem{I98}  
N. Lo Iudice, Phys.\ Rev.\ C{\bf 57},1246 (1998).

\bibitem{SH-exp}  
M. Schmidt and H. Haberland, private communication.

\bibitem{FP96}  
S. Frauendorf and V.V. Pashkevich, Ann.\ Phys. (Leipzig) {\bf 5}, 34 (1996).

\bibitem{C85}  
K. Clemenger, Phys.\ Rev. B {\bf 32}, 1359 (1985).
\end{references}
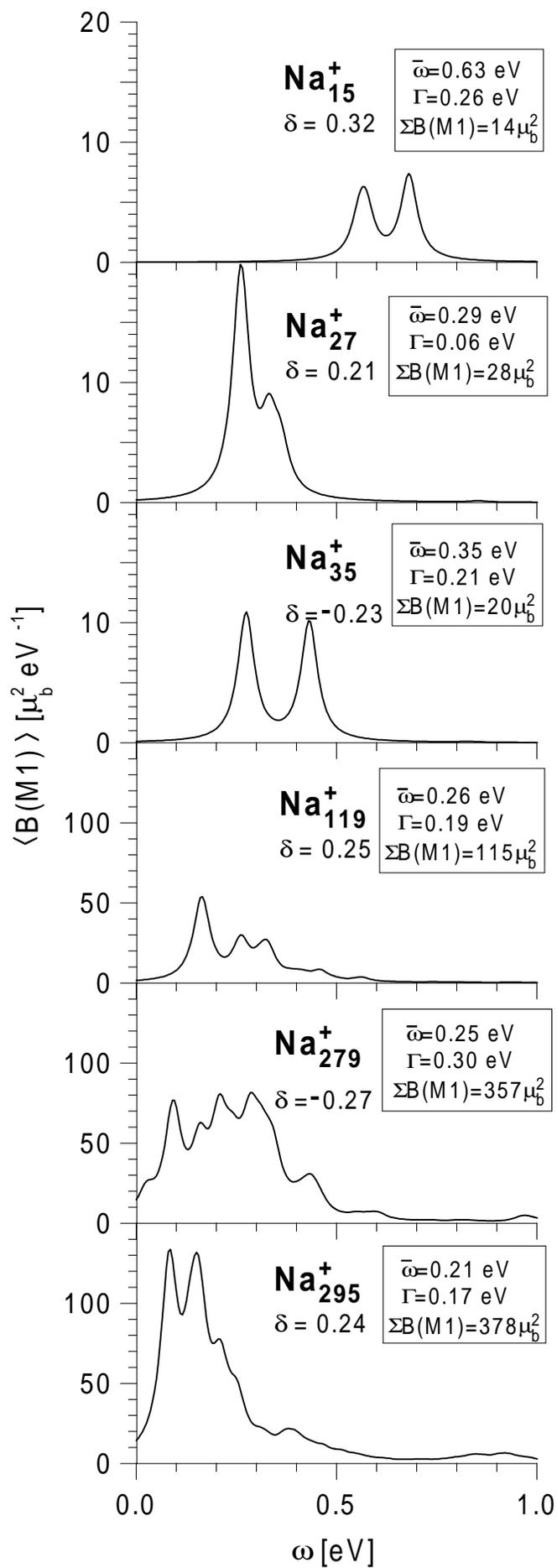
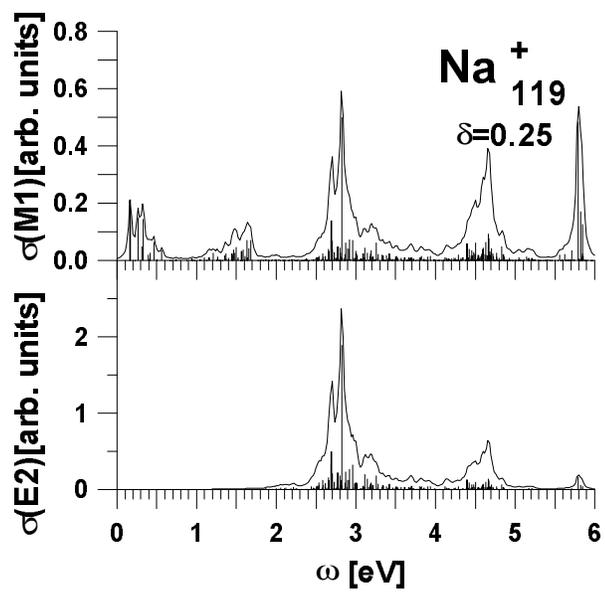
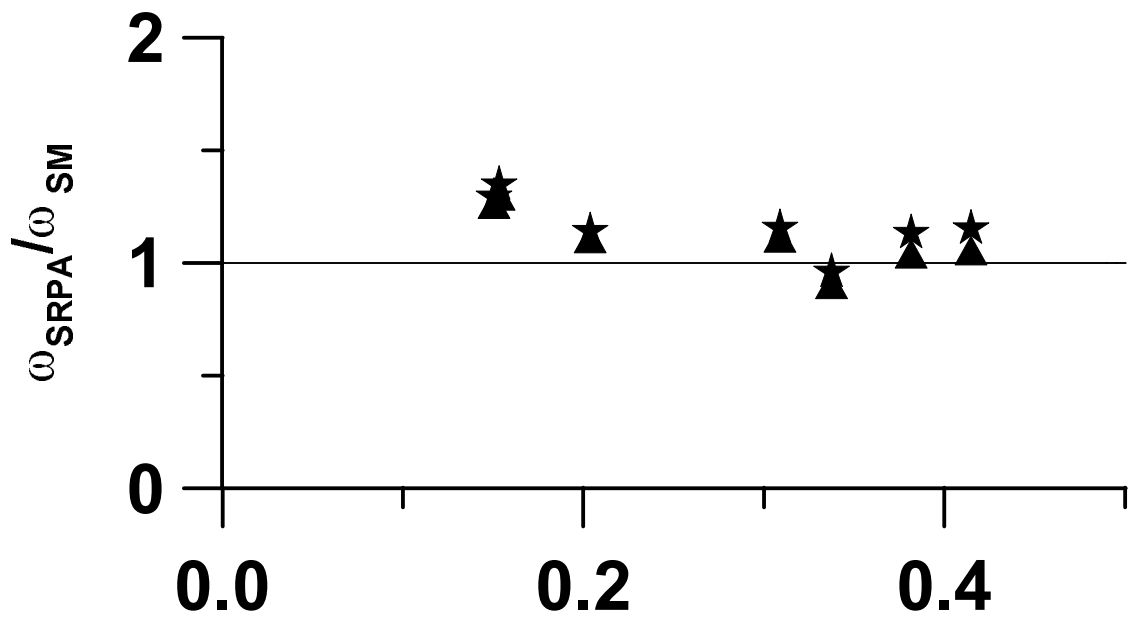
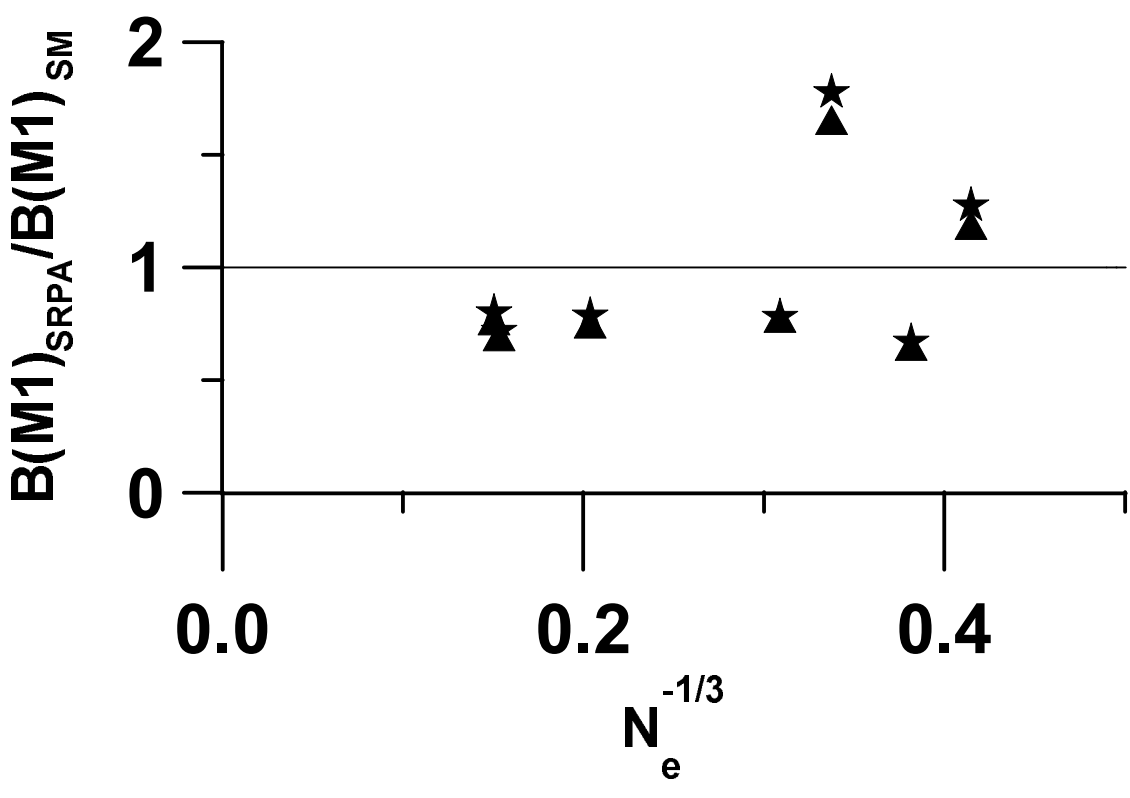
\end{document}